\documentclass[11pt]{article}
\usepackage{geometry}  
\usepackage[T1]{fontenc}  
\usepackage[utf8x]{inputenc}  
\usepackage{lineno}  
\usepackage{fancyhdr}  
\usepackage{enumitem}  
\usepackage{hyperref}  
\usepackage{titling}  
\usepackage[numbers]{natbib}  
\usepackage{mathtools}  
\usepackage{titlesec}  
\usepackage{lastpage}  
\usepackage[english]{babel}  
\usepackage{amsmath}  
\usepackage{amsfonts}  
\usepackage{amssymb}  
\usepackage{wasysym}  
\usepackage{bbm}  
\usepackage{array}  
\usepackage{xr}  
\usepackage{verbatim} 
\usepackage{float} 
\usepackage{color}
\usepackage{algorithm, caption}
\usepackage[noend]{algpseudocode}
\usepackage{xpatch} 
\usepackage{xspace} 
\usepackage{booktabs}
\usepackage[table,xcdraw]{xcolor}
\usepackage{tabularx}
\usepackage{tabulary}
\usepackage{multirow}
\usepackage{comment}

\usepackage{subcaption}
\usepackage{scalerel}

\usepackage[normalem]{ulem}
\usepackage{soul}

\usepackage{graphicx}

\hypersetup{%
  pdfborderstyle={/S/U/W 1}
}

\setcitestyle{round,numbers}
\makeatletter
\def\BState{\State\hskip-\ALG@thistlm}
\makeatother

\newcommand{\email}[1]{\href{mailto:{#1}}{{#1}}}

\newcommand{\keywords}[1]{\textbf{Keywords}: {#1}}

\newcommand{\optincludegraphics}[2][]{}

\newcommand{\optinput}[1]{}

\newtagform{brackets}{[}{]}
\usetagform{brackets}

\title{Technical Report (v1.0)--Pseudo-random Cartesian Sampling for Dynamic MRI}

\immediate\write18{texcount -sum=1,1,1,1,1,1,1 -merge -incbib -dir -utf8 \jobname.tex > \jobname.wcTotal }
\immediate\write18{texcount -sub=none -merge -incbib -dir -utf8 \jobname.tex | grep _main_ | grep -oE '[0-9]+' | python -c "import sys; print(sum(int(l) * w for l, w in zip(sys.stdin, [1, 1, 0, 0, 0, 0, 0])))" > \jobname.wcManuscript }
\immediate\write18{texcount -sub=none -merge -incbib -dir -utf8 \jobname.tex | grep Abstract | grep -oE '[0-9]+' | python -c "import sys; print(sum(int(l) * w for l, w in zip(sys.stdin, [1, 1, 0, 0, 0, 0, 0])))" > \jobname.wcAbstract }

\newcommand{\wcTotal}{\clearpage{\noindent\large{\bf Detailed Word Count} (not to be included for submission)}\verbatiminput{\jobname.wcTotal}}
\newcommand{\wcManuscript}{\input{\jobname.wcManuscript}}
\newcommand{\wcAbstract}{\input{\jobname.wcAbstract}}

\begin{document}

    
\begin{titlepage}
\begin{center}
{\noindent\LARGE\bf \thetitle}

\bigskip
Date Modified: \today

\bigskip
\bigskip
\large
    Mihir Joshi,\textsuperscript{1}
    Aaron Pruitt,\textsuperscript{2}
    Chong Chen,\textsuperscript{1}
	Yingmin Liu,\textsuperscript{3}
	Rizwan Ahmad,\textsuperscript{*1,3,4}
\end{center}

\bigskip

\noindent
\begin{enumerate}[label=\textbf{\arabic*}]
\item Biomedical Engineering, The Ohio State University, Columbus OH, USA
\item Siemens Healthineers, Chicago, IL, USA
\item Davis Heart \& Lung Research Institute, The Ohio State University, Columbus OH, USA
\item Electrical and Computer Engineering, The Ohio State University, Columbus OH, USA
\end{enumerate}

\bigskip
\bigskip


\begin{flushleft}
\textbf{*} Corresponding author:

\begin{tabular}{>{\bfseries}rl}
Name		& Rizwan Ahmad													\\
Department	& Biomedical Engineering													\\
Institute	& The Ohio State University														\\
Address 	& 460 W 12th Ave, Room 318														\\
			& Columbus OH 43210, USA														\\
E-mail		& \email{ahmad.46@osu.edu}											\\
\end{tabular}
\end{flushleft}

\bigskip

\begin{flushleft}
This work was partially unfunded by NIH grants R01HL135489 and R01HL151697.											\\
\end{flushleft}

\vfill




\end{titlepage}

\pagebreak

\begin{abstract}

\noindent

For an effective application of compressed sensing (CS), which exploits the underlying compressibility of an image, one of the requirements is that the undersampling artifact be incoherent (noise-like) in the sparsifying transform domain. For cardiovascular MRI (CMR), several pseudo-random sampling methods have been proposed that yield a high level of incoherence.
In this technical report, we present a collection of five pseudo-random Cartesian sampling methods that can be applied to 2D cine and flow, 3D volumetric cine, and 4D flow imaging. Four out of the five presented methods yield fast computation for on-the-fly generation of the sampling mask, without the need to create and store pre-computed look-up tables. In addition, the sampling distribution is parameterized, providing control over the sampling density. For each sampling method in the report, (i) we briefly describe the methodology, (ii) list default values of the pertinent parameters, and (iii) provide a publicly available MATLAB implementation. 


\end{abstract}
\bigskip

\pagebreak

\section{Highlights of the Sampling Methods}
This technical report is not a survey or review of various sampling methods for dynamic MRI. Our aim is to summarize five Cartesian sampling methods for dynamic MRI that have been developed at our institute.  In this section, we highlight the main features of these sampling methods. Here, the term ``sample'' refers to a readout line along the $k_x$ direction. A more detailed technical description of each method is provided in the subsequent sections. A unified MATLAB implementation of these methods can be found here: \url{https://github.com/MihirJoe/cmr-sampling}. To start, execute {\sf main.m} file in MATLAB.

\subsection{VISTA Highlights}
VISTA stands for Variable density Incoherent SpatioTemporal Acquisition. VISTA performs sampling in the $k_y$-$t$ domain.
\begin{enumerate}
    \itemsep0em
    \item The sampling is based on Fekete points \cite{bendito2007estimation}.
    \item Previous download link: \url{https://github.com/OSU-CMR/VISTA}
    \item Related publication \cite{ahmad2015variable}: \url{https://onlinelibrary.wiley.com/doi/abs/10.1002/mrm.25507}
    \item Applications: 2D dynamic MRI, e.g., cardiac cine and first-pass perfusion.
    \item Notes: VISTA is computationally slow, especially for large problems. For clinical real-time cine, it has been replaced with Golden Ratio Offset sampling (discussed below) at our institute.
\end{enumerate}

\subsection{GRO Highlights}
GRO stands for Golden Ratio Offset sampling. It performs sampling in the $k_y$-$t$-encoding domain. 
\begin{enumerate}
\itemsep0em
    \item The sampling is based on golden ratio shifts between adjacent frames.
    \item Previous download link: \url{https://github.com/OSU-CMR/GRO-CAVA}
    \item Related publication: Apart from a patent application, this method has not appeared in a peer-reviewed publication \cite{ahmad2022cartesian}. A brief description of the method is provided in Section~\ref{sec:gro}. Cite this technical report if you use GRO as a sampling pattern in your work.
    \item Applications: 2D dynamic MRI, e.g., cardiac cine, phase-contrast MRI, first-pass perfusion.
    \item Notes: GRO is preferred over VISTA due to fast computation speed. Also, GRO supports sampling for phase-contrast MRI, with multiple velocity encoding directions. Due to smaller k-space jumps between subsequent samples, both VISTA and GRO are suitable for steady-state free precession (SSFP) based sequences.
\end{enumerate}

\subsection{CAVA Highlights}
CAVA stands for CArtesian sampling with Variable density and Adjustable temporal resolution. It performs sampling in the $k_y$-$t$-encoding domain.
\begin{enumerate}
\itemsep0em
    \item The sampling is based on golden ratio offset from one sample to the next.
    \item Previous download link: \url{https://github.com/OSU-CMR/GRO-CAVA}
    \item Related publication: \url{https://onlinelibrary.wiley.com/doi/abs/10.1002/mrm.28059}
    \item Applications: 2D dynamic MRI, e.g., cardiac cine and phase-contrast MRI.
    \item Notes: The acquisition based on CAVA allows retrospective adjustment of temporal resolution at the cost of marginally sacrificed uniformity compared to GRO. Due to potentially large k-space jumps, this pattern may not be suitable for SSFP-based sequences \cite{bieri2005analysis}.
\end{enumerate}

\subsection{OPRA Highlights}
OPRA stands for Ordered Pseudo RAdial sampling. It performs sampling in the $k_y$-$k_z$-$t$ domain.
\begin{enumerate}
\itemsep0em
    \item OPRA pseudo-randomly samples ``L''-shape leaflets in $k_y$-$k_z$. Each subsequent leaflet is then rotated by a golden angle with respect to the previous leaflet.
    \item Related publication: This method has not appeared in a peer-reviewed publication. A brief description of the method is provided in Section 3.  Cite this technical report if you use OPRA as a sampling pattern in your work.  
    \item Applications: 3D dynamic MRI, e.g., 3D cine, 3D perfusion, and dynamic MR angiography
    \item Notes: Due to smaller k-space jumps between subsequent samples, OPRA is suitable for SSFP-based sequences.
\end{enumerate}

\subsection{PR4D Highlights}
PR4D stands for Pseudo Radial sampling for 4D flow. It performs sampling in the $k_y$-$k_z$-$t$-encoding domain.
\begin{enumerate}
\itemsep0em
    \item Sampling technique for flow imaging defined in the polar coordinate space. 
    \item Related publication: \url{https://onlinelibrary.wiley.com/doi/full/10.1002/mrm.28491}  
    \item Applications: 3D dynamic MRI, e.g., 3D cine, 3D perfusion, dynamic MR angiography, and 4D flow
    \item Notes: Due to potentially larger k-space jumps between subsequent samples, PR4D may not be suitable for SSFP-based sequences.
\end{enumerate}



\section{VISTA Description}


\subsection{Background}
For 2D dynamic applications, we have proposed a sampling method called VISTA \cite{ahmad2015variable}. It offers many advantages over other pseudo-random Cartesian sampling methods, including (i) incorporating variable density, (ii) ensuring that the time-averaged data is fully sampled, and (iii) limiting eddy currents by controlling the extent of jumps (in k-space) from one readout to the next. VISTA is based on minimizing Riesz energy (RE) of the samples on a $k_y$-$t$ grid. RE is minimized iteratively, leading to longer computation times.

\subsection{Methodology}
VISTA is computed by solving the following optimization problem:
\begin{align}
 \boldsymbol{v} = \mathop{\textrm{argmin}}_{\boldsymbol{v}} {\frac{1}{2}\sum_{i=1}^{M} \sum_{j≠i} \frac{c(\boldsymbol{v}(i))c(\boldsymbol{v}(j))}{\|\boldsymbol{v}(i) - \boldsymbol{v}(j)\|^\beta_w}}\textrm{, subject to } \boldsymbol{v} \in C \textrm{ with } \beta > 0
 \label{eq:vista-cost}
\end{align}
where $\boldsymbol{v} = \left[\boldsymbol{v}(1),\dots,\boldsymbol{v}(M)\right]$, with $\boldsymbol{v}(i) = [k_y(i), t(i)]^{\sf{T}} $ defining the Cartesian coordinates of the $i^{\text{th}}$ sample on the $k_y$-$t$ grid. Here, $M=n\times F$ is the total number of readouts (samples), with $n$ being the number of samples per frame and $F$ being the number of frames. Also, $C$ represents a user-defined constraint that forces the number of samples in each frame to be the same and $\|\boldsymbol{v}(i)\|^\beta_{w} = (\boldsymbol{v}(i)^{\sf{T}}\boldsymbol{W}\boldsymbol{v}(i))^{\frac{\beta}{2}}$, where $\boldsymbol{W}$ is a $2\times 2$ diagonal matrix with $\boldsymbol{W}(1,1)=1$ and $\boldsymbol{W}(2,2)=w$ defining the relative scaling of the time axis with respect to the $k_y$ axis. To enforce variable density,  $c(\boldsymbol{v}(i))$ is defined to be a Gaussian function, i.e., 

\begin{align}
c(\boldsymbol{v}(i)) = c(k_y(i),t(i)) = 1 - \log_{10}(s) e^{-\frac{\left(k_y(i)-\frac{N}{2}-1\right)^2}{2\sigma^2}}
\end{align}

where $N$ is the size of the phase-encoding (PE) grid and parameters $1 \leq s \leq 10$ and $\sigma > 0$ control the sampling density profile. Note, $c(\boldsymbol{v}(i))$ is only a function of $k_y(i)$ and not $t(i)$ since sampling density variation over time is not desired. The non-convex cost function in Eq. [\ref{eq:vista-cost}] can be minimized via gradient descent. 

\begin{figure}[ht]
    \centering
    \subfloat{{\includegraphics[height=5.5cm]{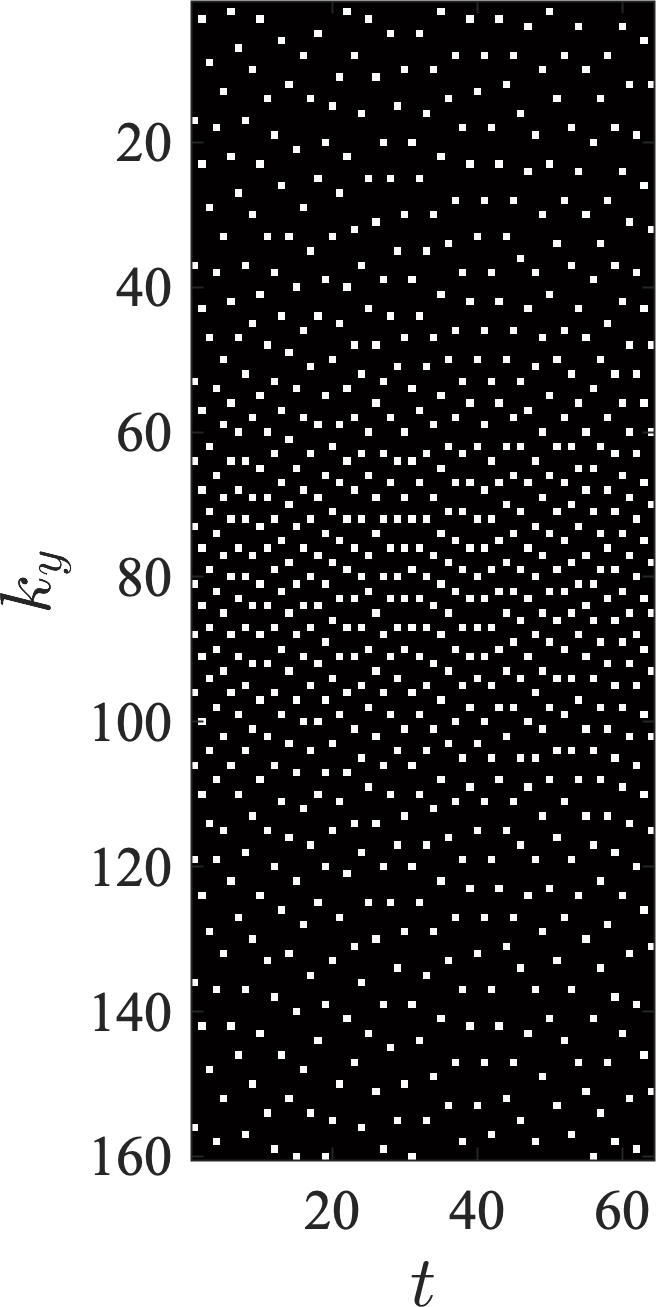} }}%
    \quad
    \subfloat{{\includegraphics[height=5.5cm]{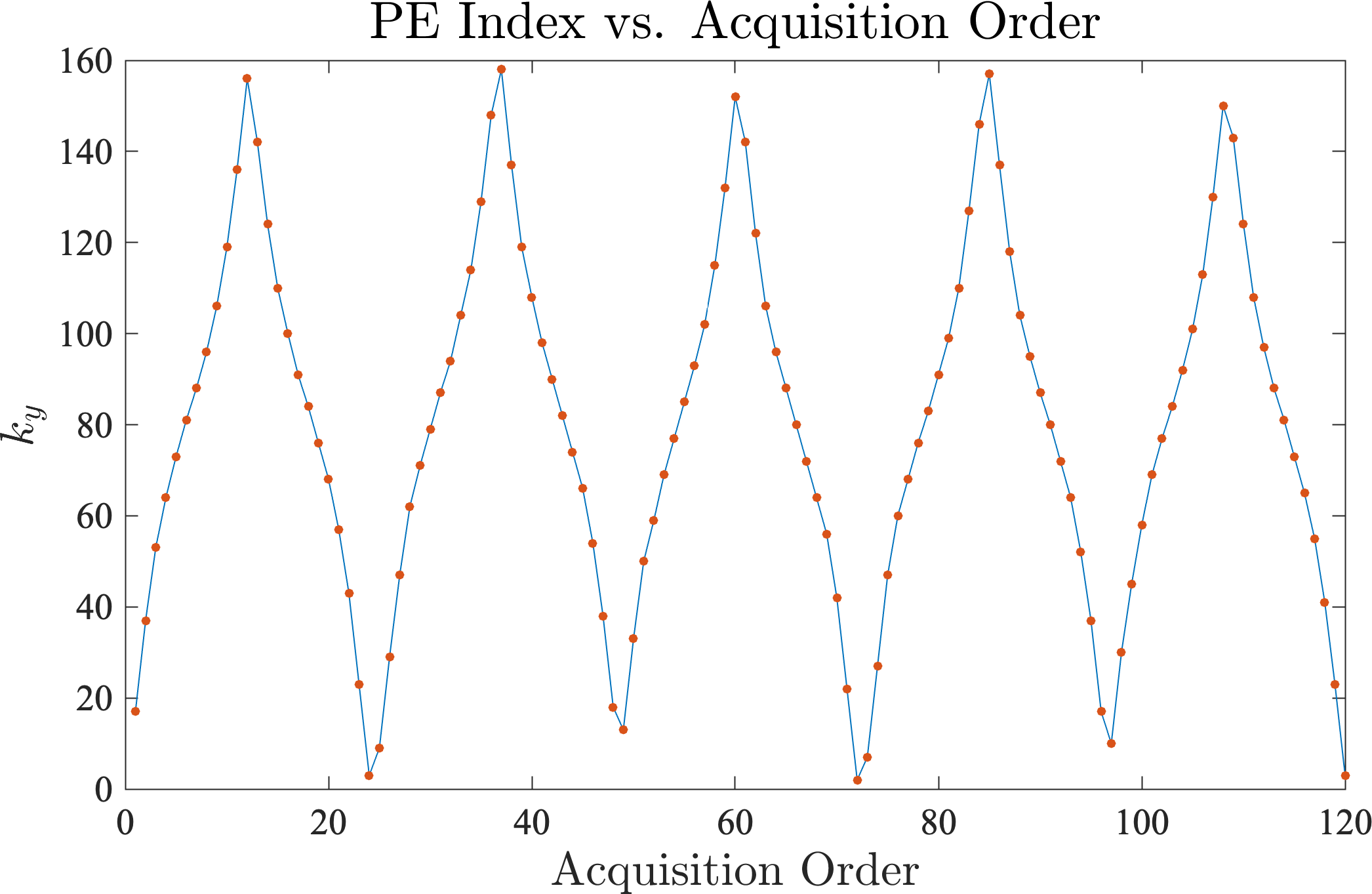} }}%
    \caption{Left: an example of VISTA sampling based on the parameter values given in Table \ref{tab:vista}. Here, $t$ represents frames and $k_y$ represents PE index. Right: PE index plotted against the order of acquisition for the first 120 samples.}%
    \label{fig:vista}%
\end{figure}

To reduce jumps in k-space, which is important for SSFP-based sequences, the PE indices in the odd frames are collected in the ascending order and the PE indices in the even frames are collected in the descending order. An example of VISTA pattern is shown in Figure \ref{fig:vista}, and the parameters used to create this pattern are reported in Table \ref{tab:vista}. 

\begin{table}[ht]
\centering
\small
\begin{tabular}{|p{2.9cm}|p{2.2cm}|l|l|}
\hline
\bf{MATLAB}  & \bf{Variables } &\bf{Description} & \bf{Default Value} \\ \hline \hline
\sf{param\_vista.PE} & $N$ & Size of PE grid & 160\\ \hline
\sf{param\_vista.FR} & $F$  & Number of frames & 64\\ \hline
\sf{param\_vista.n} & $n$ & Number of lines per frame  & 12\\ \hline
\sf{param\_vista.M} & $M$ & Total number of samples & $n\times F = 768$\\ \hline
\sf{param\_vista.s} & $s$ & Extent of variable density  & 1.6\\ \hline
\sf{param\_vista.sig} & $\sigma$ & Width of high-density region  & $N/6$ \\ \hline
\sf{param\_vista.w} & $w$ & Time dimension scaling  & $\max{\left[N/(10\times n)+0.25,1\right]}$ \\ \hline
\sf{param\_vista.beta} & $\beta$ & Norm exponent  & $1.4$ \\ \hline
\end{tabular}
\caption{Explanation of various parameters in the MATLAB implementation of VISTA.}
\label{tab:vista}
\end{table}



\section{GRO Description}
\label{sec:gro}
\subsection{Background}
Computation of VISTA is slow, and its clinical application requires pre-computing look-up tables. In response, we recently proposed another sampling method, which, in addition to offering the above-mentioned benefits of VISTA, is computationally efficient and can be extended to multiple velocity encodings \cite{ahmad2022cartesian}. The proposed method, called Golden Ratio Offset sampling (GRO), is capable of generating a large number of samples in a fraction of a second, making it suitable for clinical application. 

\subsection{Methodology}
Generating GRO sampling on a $N\times F$ grid, with $N$ being the size of the grid along the PE direction and $F$ being the number of frames, involves following steps:
\begin{enumerate}
    \item For the first frame, distribute $n$ samples uniformly on a grid of size $N_s\times 1$, with $N_s = \text{round}(N/s)$. Here, $s$ controls the acceleration rate at the center of k-space compared to the overall acceleration rate, with $s>1$ ensuring that the sampling density is higher at the center of k-space
    \label{step1}
    \item For the next frame, circularly rotate the sampling pattern in the previous frame by $\tilde{g}N_s$, where $\tilde{g} = 1/(g+\tau-1)$ with $g=(1+\sqrt{5})/2$ and $\tau$ being a user-defined positive integer.
    \label{step2}
    \item Repeat Step~\ref{step2} for all the remaining frames.
    \item To create variable density, project all samples from the $N_s\times F$ grid to the larger, final $N\times F$ by nonlinear stretching operation, where an $i^{\text{th}}$ point with PE index, $k_{y,s}(i)$, on the $N_s\times F$ grid is moved to PE index, $k_{y}(i)$, on the $N\times F$ grid, i.e.,
    \begin{equation}
        k_y(i) = \left[k_{y,s}(i)-\kappa\,\text{sign}\left(\frac{N_s}{2}-k_{y,s}(i) \right)\,\left|\frac{N_s}{2}-k_{y,s}(i) \right|^\alpha + \frac{1}{2}(N-N_s )\right],
        \label{eq:gro}
    \end{equation}
where $\alpha>0$ controls the transition from high-density central region to low-density outer region and $\left[\cdot\right]$ represents the rounding operation for odd values of $N$ and rounding-up operation for even values of $N$. The constant $\kappa$ is selected such samples $k_{y,s}(i)=1$ and $k_{y,s}(i)=N_s$ are stretched to $k_{y=1}(i)$ and $k_{y}(i)=N$, respectively.
\end{enumerate}

\begin{figure}[ht]
    \centering
    \subfloat{{\includegraphics[height=5.5cm]{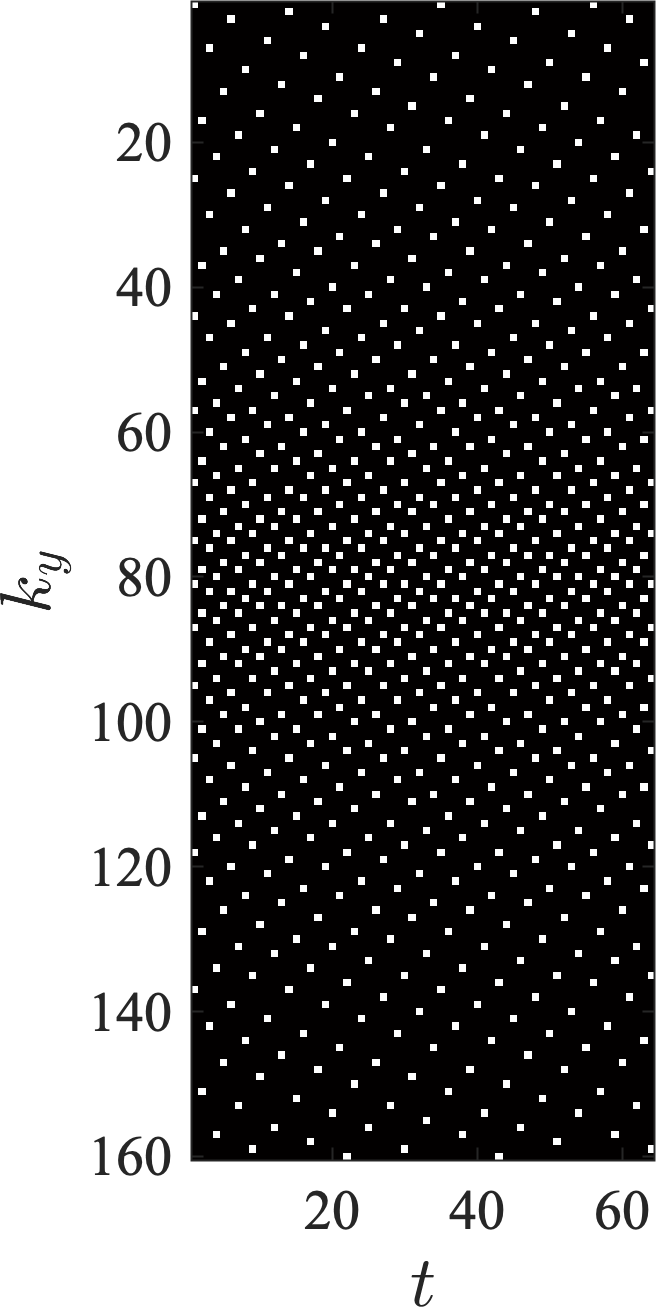} }}%
    \quad
    \subfloat{{\includegraphics[height=5.5cm]{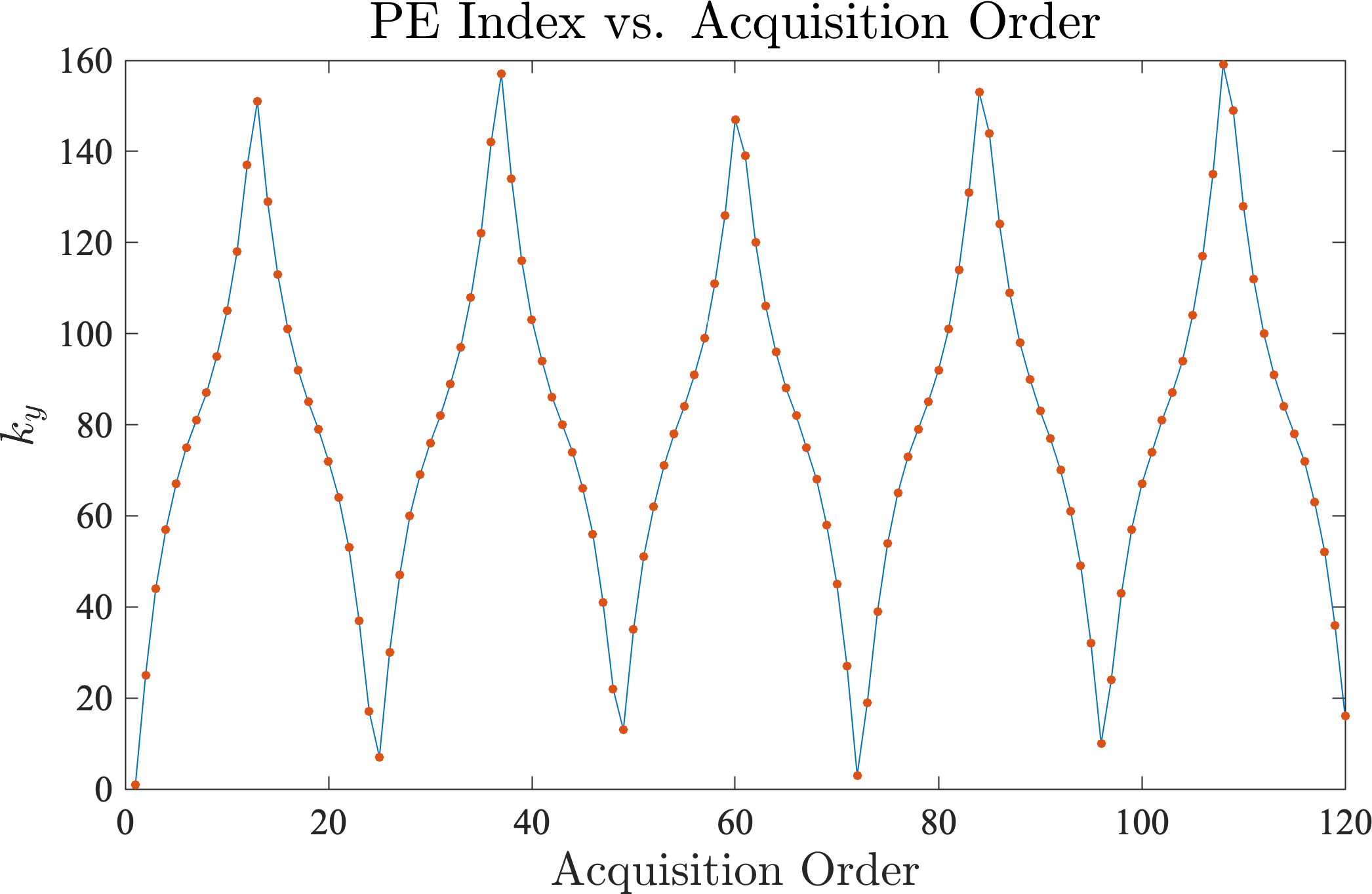} }}%
    \caption{Left: an example of GRO sampling based on the parameter values given in Table \ref{tab:gro}. Here, $t$ represents frames and $k_y$ represents phase encoding index. Right: PE index plotted against the order of acquisition for the first 120 samples.}%
    \label{fig:gro}%
\end{figure}

To reduce the size of jumps in k-space, the PE indices in the odd frames are collected in the ascending order and the PE indices in the even frames are collected in the descending order. An example of GRO pattern is shown in Figure \ref{fig:gro}, and the parameters used to create this pattern are reported in Table \ref{tab:gro}. 
By creating multiple patterns, each from a slightly shifted placement of the samples in the first frame (Step 1), GRO can be extended to phase-contrast MRI with $E>1$ encodings.   

\begin{table}[ht]
\centering
\small
\begin{tabular}{|p{2.6cm}|p{2.2cm}|l|l|l|l|}
\hline
\bf{MATLAB}  & \bf{Variables} &\bf{Description} & \bf{Default Value} \\ \hline \hline
\sf{param\_gro.PE} & $N$ & Size of PE grid  & 160\\ \hline
\sf{param\_gro.FR} & $F$ & Number of frames & 64 \\ \hline
\sf{param\_gro.n} & $n$ & Number of lines per frame & 12\\ \hline
\sf{param\_gro.M} & $M$ & Total number of samples & $n\times F = 768$\\ \hline
\sf{param\_gro.E} & $E$ & Number of encodings & 1\\ \hline
\sf{param\_gro.s} & $s$ & Extent of variable density & 2.2\\ \hline
\sf{param\_gro.alph} & $\alpha$ &  Width of high-density region & 3\\ \hline
\sf{param\_gro.tau} & $\tau$ &  Golden angle or tiny golden angle & 1\\ \hline
\end{tabular}
\caption{Explanation of various parameters in the MATLAB implementation of GRO.}
\label{tab:gro}
\end{table}

\section{CAVA Description}
\subsection{Background}
Data with most Cartesian sampling patterns, including VISTA and GRO, are collected with a predetermined temporal resolution, which cannot be changed post-acquisition. In this section, we describe CAVA \cite{rich2020cartesian}. A distinguishing features of CAVA is that the temporal resolution can be adjusted retrospectively.

\subsection{Methodology}
The first step in generating CAVA pattern on a $N\times F$ grid is creating a pseudo-random sequence of $M$ indices on a smaller grid of size $N_s = \text{round} (N/s)$. Here, $s$ controls the acceleration rate at the center of k-space compared to the overall acceleration rate, with $s>1$ ensuring that the sampling density is higher at the center of k-space. 

Starting with a randomly selected PE index, $k_{y,s}(1)$, the PE indices are advanced sequentially by $\tilde{g}N_s$, where $\tilde{g} = 1/(g+\tau-1)$ with $g=(1+\sqrt{5})/2$ and $\tau$ being a user-defined positive integer, yielding
\begin{align}
k_{y,s}(i+1)=\langle{k_{y,s}(i)+\tilde{g}N_s}\rangle_{N_s}.
\end{align}
Here, $k_{y,s}(i)$ is the PE index of the $i^{\text{th}}$ sample on the grid that has size $N_s$, and $⟨⋅⟩_{N_s}$ is the modulo operator with respect to $N_s$. To create variable density, the samples, $k_{y,s}(i)$, are stretched non-linearly and mapped to a larger grid of size $N$, yielding,
\begin{align}
k_y(i) = \left[{k_{y,s}(i)-\kappa\,\text{sign}\left(\frac{N_s}{2} - k_{y,s}(i)\right)\left|\frac{N_s}{2}-k_{y,s}(i)\right|^\alpha + \frac{1}{2}(N - N_s)}\right],
\end{align}
where $\alpha>0$ controls the transition from high-density central region to low-density outer region, and $\left[\cdot\right]$ represents the rounding operation for odd values of $N$ and rounding-up operation for even values of $N$. The constant $\kappa$ is selected such samples $k_{y,s}(i)=1$ and $k_{y,s}(i)=N_s$ are stretched to $k_{y=1}(i)$ and $k_{y}(i)=N$, respectively. 

\begin{figure}[ht]
    \centering
    \subfloat{{\includegraphics[height=5.5cm]{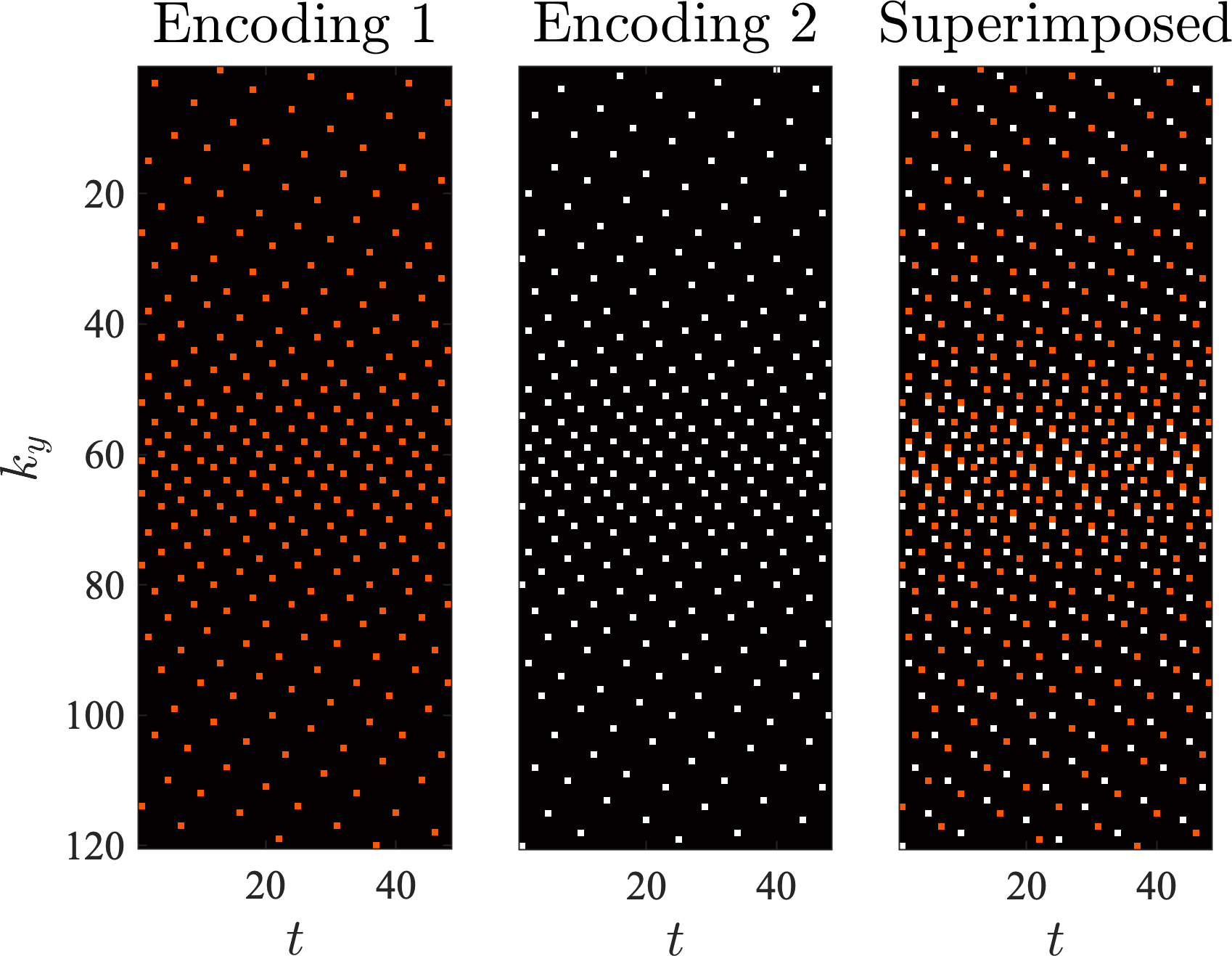} }}%
    \quad
    \subfloat{{\includegraphics[height=5.5cm]{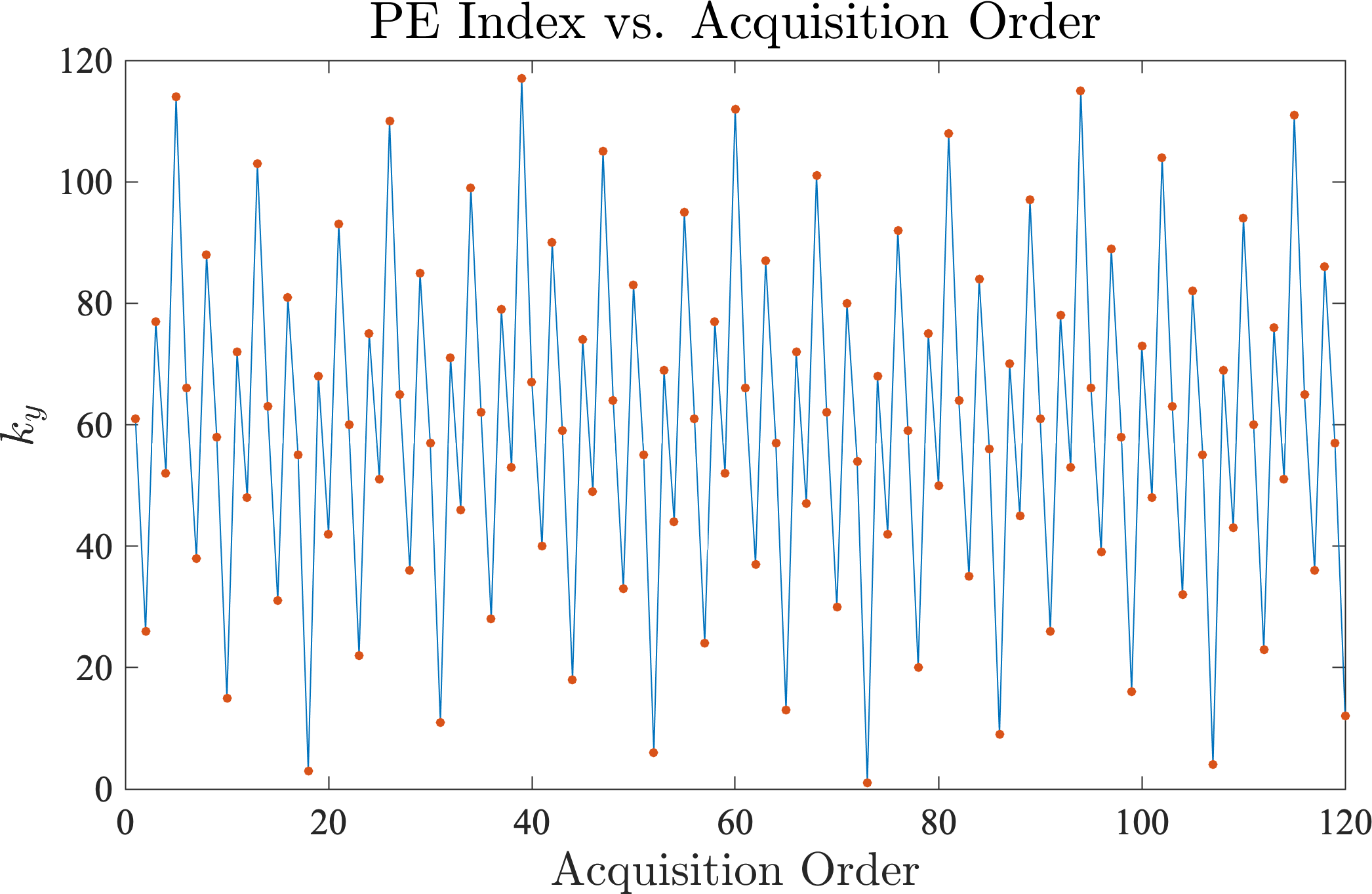} }}%
    \caption{Left: an example of CAVA sampling based on the parameter values given in Table \ref{tab:cava}. Here, $t$ represents frames and $k_y$ represents PE index. Right: PE index plotted against the order of acquisition for the first 120 samples from the first encoding.}%
    \label{fig:cava}%
\end{figure}

Due to the golden ratio jumps in k-space, CAVA sampling patterns can be re-binned to generate different temporal resolutions without sacrificing sampling uniformity within each frame. By creating multiple CAVA patterns, each from a slightly different starting point ($k_{y,s}(1)$), this sampling pattern can be extended to phase-contrast MRI with $E>1$ encodings. An example of CAVA pattern is shown in Figure \ref{fig:cava}, and the parameters used to create this pattern are given in Table \ref{tab:cava}.

\begin{table}[ht]
\centering
\small
\begin{tabular}{|p{2.9cm}|p{2.2cm}|l|l|l|l|}
\hline
\bf{MATLAB}  & \bf{Variables} &\bf{Description} & \bf{Default Value} \\ \hline \hline
\sf{param\_cava.PE} & $N$ & Size of PE grid  & 120\\ \hline
\sf{param\_cava.FR} & $F$ & *Nominal number of frames & 48\\ \hline
\sf{param\_cava.n} & $n$ & *Nominal number of lines per frame & 6\\ \hline
\sf{param\_cava.M} & $M$ & Total number of samples/endcoding & $n\times F = 288$\\ \hline
\sf{param\_cava.E} & $E$ & Number of encodings & 2\\ \hline
\sf{param\_cava.s} & $s$ & Extent of variable density & 2.2\\ \hline
\sf{param\_cava.alph} & $\alpha$ &  Width of high-density region & 3\\ \hline
\sf{param\_cava.tau} & $\tau$ &  Golden angle or tiny golden angle  & 1\\ \hline
\end{tabular}
\caption{Explanation of various parameters in the MATLAB implementation of CAVA. *The number of lines per frame (temporal resolution) and the number of frames can be changed retrospectively.}
\label{tab:cava}
\end{table}

\section{OPRA Description}

\subsection{Background}
The idea behind OPRA is to sequentially sample the $k_y$-$k_z$ grid such that the samples are evenly distributed across the entire 2D grid for any time window. This repeated coverage of $k_y$-$k_z$ allows the data to be retrospectively binned into cardiac and respiratory bins. To avoid large jumps in k-space, the samples of OPRA are restricted to reside on ``L''-shaped leaflets.


\subsection{Methodology}
To sequentially sample $k_y$-$k_z$ grid of size $N_y \times N_z$, we first place $L$ samples on an L-shaped leaflet, with elbow at $k_y=k_z=0$. To achieve variable density, the samples are more closely spaced near the center of the k-space. To cover the entire $k_y$-$k_z$ grid, the leaflets are rotated using golden angle, $\pi/g$, where $g=(1+\sqrt{5})/2$. The location of the samples on the leaflet is radially shifted, by another irrational number $g_s$, from one leaflet to the next to make sure that the resulting pattern stays incoherent. Here we describe how to place $L$ samples on a leaflet. 

Let $\tilde{\theta}(l,j)$ defines the angle (orientation) of the $j^{\text{th}}$ sample on the $l^{\text{th}}$ leaflet, 

$$
\tilde{\theta}(l,j) = \begin{cases}
\langle{(l-1)\pi/g}\rangle_{2\pi} & j=1,2,\dots,\tilde{L}\\
\langle{l\pi/g - \phi}\rangle_{2\pi} & j=\tilde{L}+1,\dots,L
\end{cases}
$$

where $l=1,2,\dots,M/L$ represents leaflets, $\tilde{L}=L/2$, and $\phi$ determines the size of k-space jump from the end of one leaflet to the next.

\begin{figure}[ht]
    \centering
    \includegraphics[height = 5.2cm]{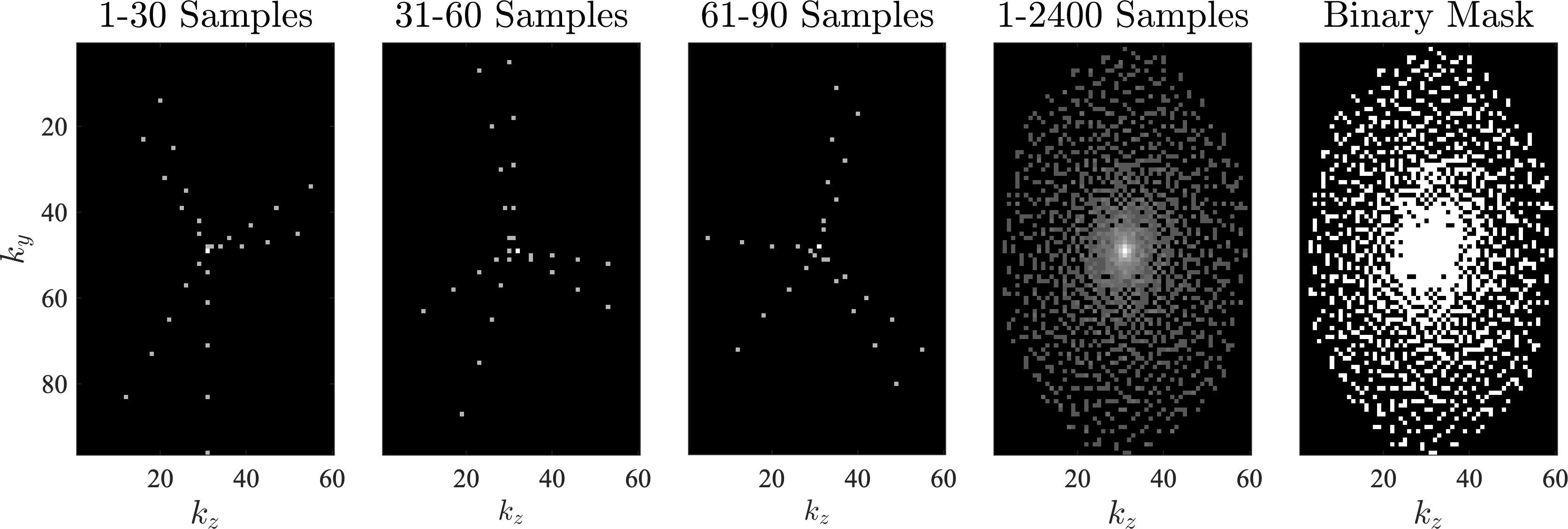}
    \caption{An example of OPRA sampling for 3D imaging. From left to right: first 30 samples (three leaflets), $31-60$ samples, $61-90$ samples, all 2400 samples superimposed, and the binary mask corresponding to the 2400 samples. }
    \label{fig:opra}
\end{figure}

For $N_y \neq N_z$, we apply the following correction to ensure that the distribution is angularly uniform
\begin{align}
\theta(l,j) = \tan^{-1}\left[\frac{\left(N_z/N_y\right)^{(\gamma+1)} \sin\left(\tilde{\theta}(l,j)\right)}{\cos\left(\tilde{\theta}(l,j)\right)} \right]
\end{align}
The maximum radii, $R(l,j)$, for the two halves of $l^{\text{th}}$ leaflet are defined as
\begin{align}
R(l,j) = \frac{N_y N_z}{\sqrt{\left(N_y \sin \left(\theta(l,j)\right)\right)^2 + \left(N_z \cos \left(\theta(l,j)\right)\right)^2}}
\end{align}
The radius of the $j^{\text{th}}$ sample on this leaflet is calculated by 
\begin{align}
\tilde{r}(l,j) = R(l,j)-(j-1)R(l,j)/\tilde{L} - \langle{(l-1)g_s}\rangle_{1}\, R(l,j)/\tilde{L},\quad j=1,2,\dots, L
\label{eq:radius}
\end{align}
The variable density is created by non-linear scaling of the radius, i.e., 
\begin{align}
r(l,j) = \tilde{r}^s(l,j)
\end{align}
 Finally, polar coordinates ($\theta(l,j)$ and $r(l,j)$) are converted to Cartesian coordinates ($k_y(i)$ and $k_z(i)$) to get the final sampling pattern, where $i = (l-1)L + j$. An
example of OPRA pattern is shown in Figure \ref{fig:opra}, and the parameters used to create this pattern are reported in Table \ref{tab:opra}.

\begin{table}[ht]
\centering
\small
\begin{tabular}{|l|l|l|l|}
\hline
\bf{MATLAB} & \bf{Variables} &\bf{Description} & \bf{Default Value} \\ \hline \hline
\sf{param\_opra.PE}  & ($N_y, N_z$) & Matrix size & (96, 60)\\ \hline
\sf{param\_opra.FR} & $F$ & *Nominal number of frames & 80\\ \hline
\sf{param\_opra.n}  & $n$ & *Nominal number of lines per frame & 30\\ \hline
\sf{param\_opra.L}  & $L$ & Number of samples in a leaflet & 10\\ \hline
\sf{param\_opra.M}  & $M$ & Total number of samples & $n\times F = 2400$\\ \hline
\sf{param\_opra.s}  & $s$ & Extent of variable density & 3\\ \hline
\sf{param\_opra.ar} & $\gamma$ & Aspect ratio of the high-density region & 0 \\ \hline
\sf{param\_opra.gs} & $g_s$ & An irrational number & $\sqrt{6}$\\ \hline
\sf{param\_opra.phi} & $\phi$ & Angular jump among leaflets & $\pi/12$\\ \hline
\end{tabular}
\caption{Explanation of various parameters in the MATLAB implementation of OPRA. *The number of lines per frame (temporal resolution) and the number of frames can be changed retrospectively.}
\label{tab:opra}
\end{table}

\section{PR4D Description}
\subsection{Background}
PR4D samples the $k_y$-$k_z$-encoding grid of size $N_y\times N_z \times E$. In contrast to OPRA, PR4D does not restrict the samples to reside on leaflets. As a result, PR4D provides more even coverage at the cost of large k-space jumps. 

\subsection{Methodology}
PR4D sampling is defined in a polar coordinate system, where radius and angle are advanced using two different irrational numbers. The angle for the $i^{\text{th}}$ sample and the $e^{\text{th}}$ encoding is defined as

\begin{align}
	\tilde{\theta}(i, e) = \left\langle2\pi g_s\left(i+\frac{e-1}{E}\right)\right\rangle_{2\pi}.  
\end{align}
Here, $i=1,2,\dots,M$ refers to the sample acquisition order with in each encoding, $e = 1, 2\dots, E$ refers to the velocity encoding index, and $g_s = \langle \sqrt[3]{35}\rangle_{1}$. To adjust the orientation of the high density region, we apply the following correction
\begin{align}
\theta(i,e) = \tan^{-1}\left[\frac{\left(N_z/N_y\right)^\gamma \sin\left(\tilde{\theta}(i,e)\right)}{\cos\left(\tilde{\theta}(i,e)\right)} \right]
\end{align}

\begin{figure}[ht]
    \centering
    \includegraphics[height = 19cm]{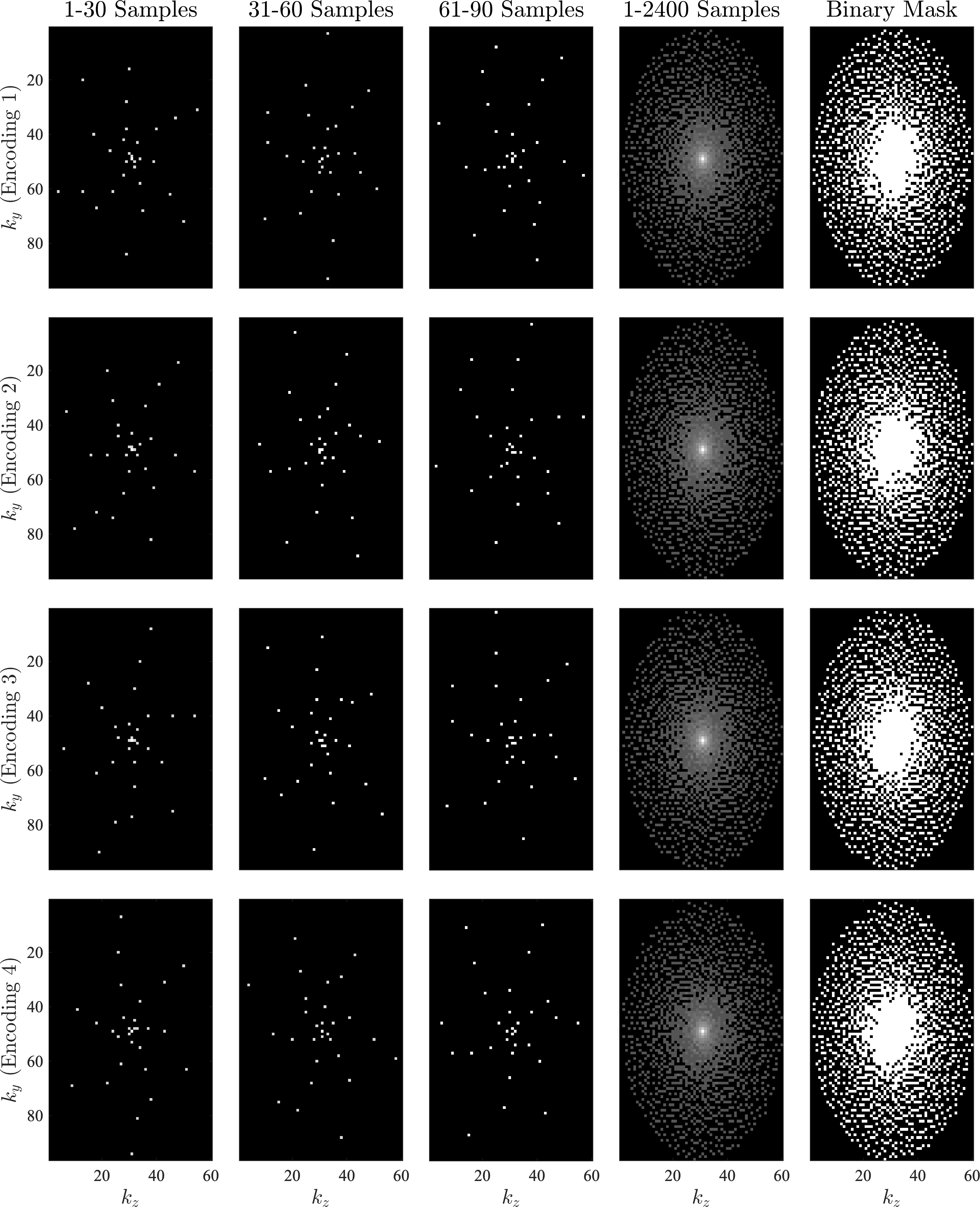}
    \caption{An example of PR4D sampling for 4D flow imaging. From left to right: first 30 samples, $31-60$ samples, $61-90$ samples, all 2400 samples superimposed, and the binary mask corresponding to the 2400 samples. The four velocity encodings are shown across rows.}
    \label{fig:pr4d}
\end{figure}

Starting from $r(1, \cdot)=0$, the radial indices are changed based on a second irrational number, $2-g$, where $g=(1+\sqrt{5})/2$, which gives 

  \begin{align}
  	 r(i+1, e) = \left[\left\langle r(i,e) + R(2-g)\right\rangle_{R}\right]^s
  	 \label{eq:nonlin-scaling}
  \end{align}
  where $R = \lfloor (\max(N_y, N_z)-1)/2 \rfloor$ represents a maximum allowed radius, with $\lfloor \cdot \rfloor$ denoting the rounding-down operation. The nonlinear scaling $\left[ \cdot \right]^s$ in Eq. [\ref{eq:nonlin-scaling}] promotes variable density. Note, $\theta$ is chosen to be a function of $e$, whereas $r$ is not. Finally, the radius is normalized and the polar coordinates ($\theta(i,e)$ and $r(i,e)$) are converted to Cartesian coordinates ($k_y(i,e)$ and $k_z(i,e)$) to get the final sampling pattern.  An example of PR4D pattern is shown in Figure \ref{fig:pr4d}, and the parameters used to create this pattern are reported in Table \ref{tab:pr4d}.

\begin{table}[ht]
\centering
\small
\begin{tabular}{|l|l|l|l|}
\hline
\bf{MATLAB} & \bf{Variables} &\bf{Description} & \bf{Default Value} \\ \hline \hline
\sf{param\_pr4d.PE}  & ($N_y, N_z$) & Matrix size & (96, 60)\\ \hline
\sf{param\_pr4d.FR} & $F$ & *Nominal number of frames & 80\\ \hline
\sf{param\_pr4d.n}  & $n$ & *Nominal number of lines per frame & 30\\ \hline
\sf{param\_pr4d.M}  & $M$ & Total number of samples & $n\times F = 2400$\\ \hline
\sf{param\_pr4d.E} & $E$ & Number of encodings & 4\\ \hline
\sf{param\_pr4d.s}  & $s$ & Extent of variable density & 3\\ \hline
\sf{param\_pr4d.ar} & $\gamma$ & Aspect ratio of the high-density region & 0 \\ \hline
\sf{param\_pr4d.gs} & $g_s$ & An irrational number & $\sqrt[3]{35}$\\ \hline
\end{tabular}
\caption{Explanation of various parameters in the MATLAB implementation of PR4D. *The number of lines per frame (temporal resolution) and the number of frames can be changed retrospectively.}
\label{tab:pr4d}
\end{table}

\clearpage







\bibliography{root.bib}
\clearpage
\end{document}